\documentclass[sigconf, screen]{acmart}

\usepackage{xspace}
\usepackage{calc}
\usepackage{multirow}
\usepackage{wrapfig}

\AtBeginDocument{%
  }

\setcopyright{acmlicensed}
\copyrightyear{2023} 
\acmYear{2023} 
\acmDOI{10.1145/3544548.3581470}

\acmConference[CHI '23]{Proceedings of the 2023 CHI Conference on Human Factors in Computing Systems}{April 23--28, 2023}{Hamburg, Germany}
\acmBooktitle{Proceedings of the 2023 CHI Conference on Human Factors in Computing Systems (CHI '23), April 23--28, 2023, Hamburg, Germany}
\acmPrice{15.00}
\acmISBN{978-1-4503-9421-5/23/04}

\acmSubmissionID{1136}

\begin{document}

\title{GeoCamera: Telling Stories in Geographic Visualizations with Camera Movements}


\author{Wenchao Li}
\orcid{0000-0003-2605-7331}
\authornote{This work was done during the internship at Microsoft Research Asia.}
\affiliation{%
  \institution{The Hong Kong University of Science and Technology}
  \city{Hong Kong SAR}
  \country{China}}
\email{wlibs@connect.ust.hk}

\author{Zhan Wang}
\orcid{0000-0003-3318-6473}
\authornotemark[1]
\affiliation{%
  \institution{The Hong Kong University of Science and Technology (Guangzhou)}
  \city{Guangzhou}
  \country{China}}
\email{zwang834@connect.hkust-gz.edu.cn}

\author{Yun Wang}
\orcid{0000-0003-0468-4043}
\authornote{Y. Wang and S. Chen are the corresponding authors.}
\affiliation{%
  \institution{Microsoft Research Asia}
  \city{Beijing}
  \country{China}}
\email{wangyun@microsoft.com}

\author{Di Weng}
\orcid{0000-0003-2712-7274}
\affiliation{%
  \institution{Microsoft Research Asia}
  \city{Beijing}
  \country{China}}
\email{diweng@microsoft.com}

\author{Liwenhan Xie}
\orcid{0000-0002-2601-6313}
\affiliation{%
  \institution{The Hong Kong University of Science and Technology}
  \city{Hong Kong SAR}
  \country{China}}
\email{lxieai@connect.ust.hk}

\author{Siming Chen}
\orcid{0000-0002-2690-3588}
\authornotemark[2]
\affiliation{%
  \institution{Fudan University}
  \city{Shanghai}
  \country{China}}
\email{siming.chen@fudan.edu.cn}

\author{Haidong Zhang}
\orcid{0000-0001-7411-8042}
\affiliation{%
  \institution{Microsoft Research Asia}
  \city{Beijing}
  \country{China}}
\email{haizhang@microsoft.com}

\author{Huamin Qu}
\orcid{0000-0002-3344-9694}
\affiliation{%
  \institution{The Hong Kong University of Science and Technology}
  \city{Hong Kong SAR}
  \country{China}}
\email{huamin@cse.ust.hk}

\renewcommand{\shortauthors}{Li et al.}

\newcommand{\tool}{GeoCamera\xspace}
\newcommand{\eg}{\textit{e.g.}}
\newcommand{\etc}{\textit{etc}}
\newcommand{\ie}{\textit{i.e.}}
\newcommand{\etal}{\textit{et al.}\xspace}
\newcommand{\side}[1]{\marginpar{\small{\color{red}{#1}}}}  
\newcommand{\says}[1]{\textit{``#1''}}

\newlength\myheight
\newlength\mydepth
\settototalheight\myheight{Xygp}
\settodepth\mydepth{Xygp}
\setlength\fboxsep{0pt}
\newcommand*\inlinegraphic[1]{%
  \settototalheight\myheight{Xygp}%
  \settodepth\mydepth{Xygp}%
  \raisebox{-0.8\mydepth}{\includegraphics[height=1.2\myheight]{#1}}%
}

\newcommand{\wenchao}[1]{\textcolor{cyan}{Wenchao: #1}}
\newcommand{\shelly}[1]{\textcolor{teal}{[Shelly: #1]}}
\newcommand{\todo}[1]{{\color{red}{[TODO: #1]}}}
\newcommand{\draft}[1]{{\color{blue}{#1}}}
\newcommand{\rev}[1]{{\color{black}{#1}}}
\newcommand{\reviewer}[1]{\textcolor{orange}{[\textbf{Reviewer:} #1]}}
\newcommand{\new}[1]{\textcolor{blue}{#1}}

\begin{abstract}
In geographic data videos, camera movements are frequently used and combined to present information from multiple perspectives. However, creating and editing camera movements requires significant time and professional skills. This work aims to lower the barrier of crafting diverse camera movements for geographic data videos. First, we analyze a corpus of 66 geographic data videos and derive a design space of camera movements with a dimension for geospatial targets and one for narrative purposes. Based on the design space, we propose a set of adaptive camera shots and further develop an interactive tool called \emph{GeoCamera}. This interactive tool allows users to flexibly design camera movements for geographic visualizations. We verify the expressiveness of our tool through case studies and evaluate its usability with a user study. The participants find that the tool facilitates the design of camera movements.


\end{abstract}

\ccsdesc[500]{Human-centered computing~Interactive systems and tools}

\keywords{Visual storytelling, data video, geographic visualization, authoring tools}


\maketitle


\section{Introduction}
\label{sec:introduction}

\begin{figure*}[th]
    \centering
    \includegraphics[width=0.98\textwidth]{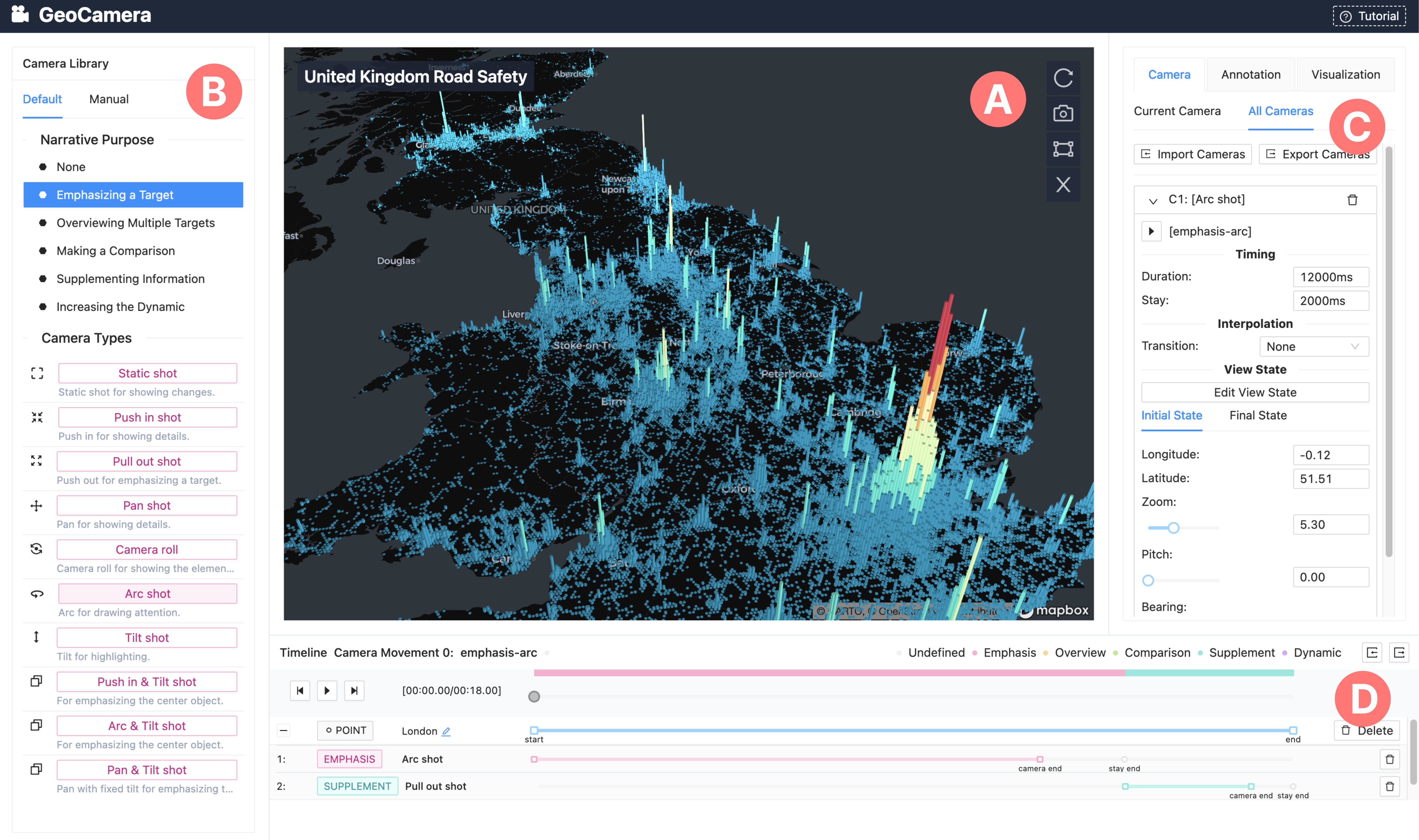}
    \caption{The user interface of \tool: (A) previewer of camera movements in the geographic visualization with geospatial targets selection, (B) library of camera shots classified by narrative purposes, (C) camera and visual configurations and crafted examples, and (D) location-camera hierarchical timeline.}
    \Description{Figure 1 describes the layout of the system interface. }
    \label{fig:ui}
\end{figure*}

Geographic data videos have been prospering for years as an intuitive storytelling medium for geographic data. In practice, geographic data videos increasingly comprise diverse camera movements (\eg, ~\cite{geovideo1,geovideo2,geovideo3}) to depict a sequence of geographic insights. The use of appropriate camera movements is essential in the authoring of geographic data videos. First, camera movements help present geographic visualizations from suitable viewing angles. For example, a 3D bar chart placed on a map needs to be read from a lower viewpoint, and the occlusions among the bars can be reduced by animating the camera around the chart. Second, these camera movements help change the narrative focus in the geographic context smoothly, which is beneficial for driving the narration~\cite{tang2020transition, cheng2022investigating}. Third, the dynamic camera movements naturally attract attention~\cite{tversky2002animation} and engage the audience~\cite{amini2015datavideo}. Finally, by changing the moving speed of the camera, certain emotions can be delivered to the audience; for example, pulling the camera very slowly away from objects to suggest their isolation and loneliness~\cite{lan2022negative}.


However, it remains challenging for people without knowledge in filmmaking to design camera movements in geographic data videos. On the one hand, although a rich palette of camera movements is available~\cite{katz1991film}, guidance in choosing the most appropriate one for a particular geographic insight or target under the story contexts is lacking; on the other hand, off-the-shelf toolkits for authoring camera movements fail to strike a balance between granularity (expressiveness) and agency (complexity in human operations)~\cite{mendez2018considering}. For instance, kelper.gl~\cite{kepler}, a geospatial analytics platform, supports the animation of the temporal changes in geospatial data points from a fixed point of view. Power Map~\cite{powermap}, a plugin for Microsoft Excel, enables users to fly between locations with a predefined camera movement on a geographic visualization. To gain more flexibility in editing camera movements, people often resort to complex video-editing software (\eg, Adobe After Effect~\cite{aftereffect}) or programming frameworks for geographic visualizations (\eg, deck.gl~\cite{deck} and Mapbox~\cite{Mapbox}). However, given the difficulty in configuring complex low-level parameters, including the longitude, latitude, focal length, camera distance, and movement duration, these approaches impose a substantial barrier for average users. Setting up many scenes by manually adjusting these parameters is laborious and time-consuming.

To facilitate the authoring of camera movements, we propose GeoCamera, a new geographic data video authoring tool that seeks balance between granularity and agency. GeoCamera is designed with two goals, namely, to produce compelling camera movements for storytelling (\textbf{G1}) and to lessen the barriers in authoring geographic data videos (\textbf{G2}).


To achieve the goals, this research first investigates how camera movements have been used in practice to tell a geographic data story from different perspectives.
We collected a corpus that includes 66 geographic data videos and 805 camera movements. By analyzing \textit{why} camera movements are employed (narrative purposes), \textit{what} objects are the focus of narration (geospatial targets), and \textit{how} camera movements are designed (camera shots), we derived a design space for the camera movements in geographic data videos. 
Based on our preliminary findings, we modeled geographic data videos as a composition of scenes and implemented \tool to facilitate the creation of camera movements driven by narrative purposes. 
We demonstrate the expressiveness of GeoCamera with a case study and a reproduction study. GeoCamera was also evaluated with a user study and received positive feedback on usability and learnability from the participants.

The contributions of this study are summarized as follows.
\begin{itemize}
    \item A design space for camera movements in geographic data videos. It comprises three dimensions: narrative purposes, geospatial targets, and camera shots.
    \item A geographic data video authoring tool that facilitates average users to narrate stories with flexible camera movements tailored for different purposes.
\end{itemize}

\section{Related Work}
\label{sec:related-work}

\subsection{Understanding Data Videos}
Data videos are short data-driven motion graphics~\cite{amini2020phd}.
It is one of the basic genres in narrative visualizations~\cite{segel2010stories}.
Researchers have been interested in understanding how to create data videos expressively and efficiently.
Prior works often carried out a qualitative examination to capture salient characteristics in existing pieces, including content analysis, user studies, or formative interviews.
Amini \etal~\cite{amini2015datavideo} analyzed 50 data videos and summarized visual representations and attention cues of data video content.
Upon the cinematic definition of four major narrative categories (\ie, \textit{establisher}, \textit{initial}, \textit{peak} and \textit{release}), they revealed common narrative structure patterns in data videos.
Shi \etal~\cite{shi2021animation} studied animation in data videos with regard to cinematography and summarized 4 animation techniques and 8 visual narrative strategies from 82 examples.
Xu \etal~\cite{xu2021cinematicopening} focused on the openings of data videos and summarized 6 types of cinematic styles out of hundreds of classic films.
Amini \etal~\cite{Amini2018hook} conducted crowdsourced studies and found that pictorial or animation representations improved viewers' engagement with data videos.

In addition to the cinematic aspects, recent works turned to narrative methodologies in data videos.
Cao \etal~\cite{cao2020examining} analyzed 70 data and presented a taxonomy, including 5 distinct genres, 4 narrative structures, and 6 narrative attributes.
Shu \etal~\cite{shu2020makes} proposed a design space for data-GIFs from 108 examples.
They further studied the impact on the understandability of each design dimension through interview and questionnaire studies.
Yang \etal~\cite{yang2022freytag} investigated 103 data videos to understand how Freytag's Pyramid, a well-received narrative structure, has been utilized.
Similarly, we applied content analysis to understand the cinematic styles and animated transitions in data videos.
Our concentration on geographic data videos revealed a fruitful subset that encompasses richer camera effects than general data videos.
In addition, unlike previous work that mostly contributed abstract design implications, our empirical findings were directly applied in our authoring tool design.

\subsection{Authoring Geographic Data Videos}
Owning to the ubiquity of geospatial data, geographic data stories have been a common category in data-driven storytelling~\cite{tong2018storytelling}, where maps are essential to provide the spatial context~\cite{caquard2013cartography, hografer2020state}.
Prior works have synthesized plenty of implications for designing a geospatial data story.
For instance, Nagel~\etal demonstrated how staging transitions could effectively explain steps from a high-level view to a fine-grained view through a public exhibition design~\cite{nagel2017staged}.
Mayr and Windhager~\cite{mayr2018spacetime} suggested how standard spatiotemporal visualization techniques affected narrative cognitive processing.
Roth~\cite{robert2021cartographic} proposed a design space of spatial narratives with three dimensions: narrative elements, genres, and tropes.
Furthermore, Latif \etal~\cite{latif2021deepunderstanding} showed the textual narrative sequence and its relationship with visual counterparts in geographic data stories.

However, we observe a barrier to crafting a geographic data video that tells a story in a comprehensive and palatable manner.
On the one hand, general-purposed visualization authoring tools provide limited support for geographic data~\cite{cheng2022investigating}.
Template-based or automatic tools (\eg, DataClip~\cite{amini2017dataclips}, Flourish Studio~\cite{flourish}, and AutoClip~\cite{shi2021autoclips}) ignored a rich palette of visual representations for geospatial data, such as 3D globe visualizations.
According to our corpus analysis, tools that feature higher expressibility (\eg, Ellipsis~\cite{satyanarayan2014ellipsis}, DataAnimator~\cite{thompson2021data}, and Animated Vega-Lite~\cite{zong2022animated}) do not consider camera design, which is common in geographic data videos.
Some tools are tailored for specific scenarios (\eg, ~\cite{tang2022smartshots,chen2021augmenting, lu2020timevideo, shin2022roslingifier}), yet none applies to geospatial data. 
On the other hand, off-the-shelf visualization software (\eg, Tableau~\cite{tableau}, ArcGIS~\cite{arcgis}, and Mapbox~\cite{Mapbox}) or library (\eg, deck.gl~\cite{deck}, kepler.gl~\cite{kepler}, and BigQuery~\cite{BigQuery}) are flexible for analytical tasks in geographic data, but they hardly address the need for storytelling or raise high barriers. 
Video makers may need to screen-record their operations and edit through general video tools or write external scripts programmatically.
GeoTime~\cite{Eccles2007geotime} was one of the earliest works that integrated storytelling into a data exploration tool.
With several primitive features, it helped capture the analysts' insights and support later communication.
The GAV toolkit~\cite{lundblad2013geovisual}) facilitated geographic storytelling within an interactive web context.
Most relevant to our focus on geographic data videos, Power Map~\cite{powermap} automatically generated map transitions among consequent slides. 
Our work summarizes representative camera movements in geographic data videos into a design space. We further develop an authoring tool based on a design space that supports inexperienced users to integrate various camera effects into their data videos by selecting their narrative purposes.
\subsection{Camera Effects in Data-driven Storytelling}
Camera effects orient from cinematography, with typical examples, such as trucking, tracking, zooming, rolling, and tilting~\cite{katz1991film, bordwell2013narration}.
With the pressing need to navigate audiences in the 3D space, camera control and motion planning have been intensively studied in fields related to computer graphics~\cite{christie2008camera}, including terrain visualization~\cite{serin2012automatic}, volume visualization~\cite{hsu2013multi, zheng2011iview}, game engines~\cite{halper2001camera}, robotics~\cite{kavraki1996probabilistic}, \rev{urban scene reconstruction~\cite{zhou2020offsite, liu2021aerial, liu2022learning}}, and \rev{virtual cinematography~\cite{he1996virtual, xie2018creating}}.
Prior research has validated camera effects as an important construct of data stories, which remain effective for narration guidance~\cite{tang2020design}, aesthetic enjoyment~\cite{shi2021animation}, and emotion delivery~\cite{lan2021kineticharts}.
Segel and Heer~\cite{segel2010stories} studied narrative visualizations and decomposed general visual narrative tactics into visual structuring, highlighting, and transition guidance.
They regarded camera motions as a strategy that offers transition guidance, and they found that camera zoom contributed to highlighting.
Amini \etal~\cite{amini2017dataclips} summarized nine major attention cues in data videos, including camera angle and zoom, that helped engage the audience.
Stolper \etal~\cite{stolper2016emerging} studied web-based data stories and identified linking separated story elements through animation as an emerging and recurring technique.
Although their corpus was based more on interactive webpages, our focus on data videos shares similarities in the continuous transition between the sequences of data stories.
Most relevant to our interest in geographic data stories, Cheng \etal{}~\cite{cheng2022investigating} examined the interplay of camera effects and narrations.
They concluded that map-based data clips extensively applied camera animations to steer audiences’ focus, especially for insights into locations and differences and background information.

From the authoring perspective, Thompson \etal{}~\cite{thompson2020understanding} listed camera as a type of graphics object in the design space of animated data graphics.
Alternation of the camera's configuration, such as its position or projection properties, results in a view change, such as panning, zooming, and rotating.
Tang \etal~\cite{tang2020transition} proposed a taxonomy of narrative transitions that classified camera motions as one of the five transition types, with subtypes, including pedestal, truck, tilt, pan, dolly, zoom, and rack focus.
However, their design spaces failed to capture the relationship between narratives and camera configurations.
\rev{In this work, we attempt to bridge the gap between narratives and camera configurations with empirical correlations. We recommend camera configurations suitable for the users' narrative goals, which alleviates users' burden in tweaking relevant parameters.}
Though prior works in other areas have contributed various methods for creating camera effects easily, such as optimizing camera trajectories~\cite{joubert2015interactive, xie2018creating}, controlling camera motions~\cite{serin2012automatic, huang2016trip}, and \rev{selecting view-points~\cite{assa2010virtual, zheng2011iview, hsu2013multi}}, our work differs from them in that we do not concern low-level details of the camera configuration rather than a high-level cinematic type because we focus on the coherence among given geographic data insights.

\section{Characterizing the Design Space}
In this section, we explain how we identified the design patterns of the camera movements in geographic data videos. Our methodology involves several steps: collecting a corpus of high-quality geographic data videos from online sources, analyzing the corpus to develop an initial design space, and validating the design space by two professional drone photographers. Next, we present our derived design space at the end of this section. 

\subsection{Data collection}
Based on the usage scenario and design considerations, we investigate the design of camera movements for geographic data videos based on real-world examples. 
To identify design principles, we first survey how existing hand-designed geo-stories fascinate audiences immersed in different insights into geographic visualizations. 
Previous research~\cite{cheng2022investigating,amini2017dataclips,shi2021understanding,tang2020transition} collected many data videos from various online resources to explore the common design patterns and performance styles of this digital storytelling.
These corpora cover a wide range of high-quality data videos. By using these corpora, we filtered 66 videos with map-based visualizations.
Although these videos may employ camera movements in map-related and map-unrelated clips, we only focus on those map-related clips in these videos.
We split these clips according to the type of camera effects and obtained 805 camera movements on the map in total. 
These examples are not comprehensive, because one data video tends to reuse the same camera movements to tell similar stories, and most of the data videos are only concerned with two-dimensional maps.
However, we aim to derive some versatile design patterns of camera movements to guide our authoring tool's implementation rather than covering all camera techniques for geographic videos. 

\subsection{Analysis and Validation}
Our target audiences are general users without any experience in crafting camera movements of videos, only their initial imagination for every story segment of the final video. 
For example, videos typically introduce the brief background at the beginning of the story and then delve further into details. 
In other words, they have their own narrative purposes for different data clips. 

To identify an appropriate taxonomy of narrative purposes in geographic data videos, we conducted a literature review from narrative visualization~\cite{segel2010stories, stolper2016emerging, bach2018narrative}, data graphics design~\cite{chevalier2016animations, thompson2020understanding, li2020improving, bach2018comics, shu2020makes, shi2021understanding}, and cinematic storytelling~\cite{cutting2016evolution, cutting2011act}.
After the literature review, we collected a set of important narrative purposes that camera movement can help describe key contents in a story. 
Taking this premise into consideration, three researchers coded each segment to analyze camera movements from the following aspects: (1) the type of movements (including details of specific parameters, \eg, speed) all cameras used; (2) geographic objects that camera movements describe; and (3) the narrative purpose of the segment based on the narration context. 
We iterated on this coding process until each segment of the camera movements could be coded consistently.
Through several iterations, we identified four \textit{geospatial targets} that camera movements mostly focus on and six \textit{narrative purposes} served by different camera movements to create stories to characterize our design space.
We also summarized eight basic \textit{camera shots} commonly used in geographic data videos to serve different \textit{narrative purposes} and \textit{geospatial targets}.

After initially extracting common patterns in geographic data videos, we invited two professional drone pilots outside the data visualization domain to validate the utility of our design space and refine the design space.
The newly introduced examples also demonstrated generalizability.
Both of them have been working in aerial filming for 3 to 4 years. 
We first let the two drone pilots present their video work and introduce our design space to them. Then, they were asked to use the design space to analyze their work. 
We observed how they applied the design space and found all of their presented work created by camera drones could be described along our three dimensions (\ie, narrative purposes, geospatial targets, and camera shots). 
Subsequently, we conducted an open-ended discussion with each expert to gain their feedback and suggestions for creating new geographic videos using our design space. Based on their advice, we modified our design space by merging related narrative purposes.
For example, merging the providing context and changing viewpoint into supplementing information. 
Redundant or uncommonly used camera shots were also removed to keep the items in the dimension succinct. 

The validation resulted in four geospatial targets and five narrative purposes, which were described in \autoref{sec:design-space}. 




\begin{figure*}[tb]
    \centering
    \includegraphics[width=\textwidth]{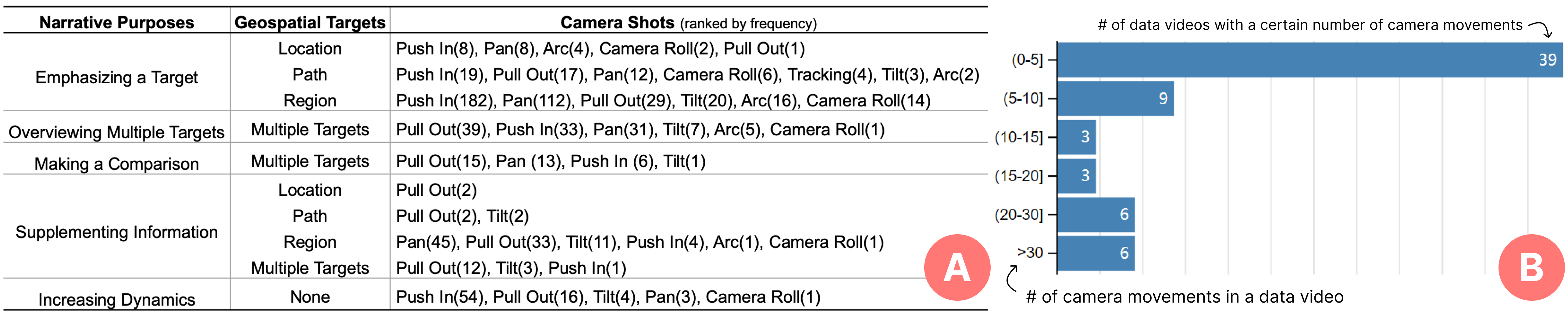}
    \caption{Statistical results of our coding on 66 geographic data videos and 805 camera movements: (A) camera shots and their corresponding narrative purposes of different geospatial targets, and (B) \rev{the distribution of camera movement number in a data video}.}
    \Description{The left part of Figure 2 uses a table to show the distribution of the camera movements in the collected geographic data videos. The right part of Figure 2 uses a histogram to show the distribution of camera movement frequency in one data video. }
    \label{fig:corpus}
\end{figure*}

\subsection{Design Space}
\label{sec:design-space}

In this section, we describe a design space of geographic data videos. The design space consists of three dimensions: narrative purposes, geospatial targets, and camera shots.
For different narrative purposes and geospatial targets, we can recommend several usable camera effects of basic camera types based on the investigation of existing geographic data videos.
\autoref{fig:corpus} shows the distribution of camera shots and the corresponding narrative purposes with geospatial targets and the frequency of the number of camera movements in one data video.


\subsubsection{Narrative Purposes}

The \textit{narrative purpose} explains \textit{why} a specific camera movement is used when crafting a geographic data video.
We summarized this taxonomy from existing literature and refined it from experts.
Notably, the same camera movement can serve multiple narrative purposes, depending on its graphic parameters and relation to geospatial targets. 
For example, a zoom-out shot is usually used for revealing the surroundings of the focal object, whereas such a shot at a slower speed commonly serves to increase the dynamic of the current scene.

\textbf{Emphasizing a Target} is a crucial criteria of narratology to make a narrative a narrative. 
Placing those most important parts at the center of the story is required~\cite{fludernik2009introduction}.
In geographic data videos, it is used to highlight geospatial targets specifically by making them occupy the largest percentage of the scene.
Emphasizing a target is the most frequently used purpose in geographic data videos.
For example, if designers want to talk about or focus on partial visualizations, then they will select an engaging way to attract audiences to partial visualizations, such as zooming in to enlarge the important parts, or by presenting such visualizations with a relatively longer duration.
When emphasizing a target, video authors sometimes adopt more combinations of different camera shots to enhance engagement compared with other narrative purposes.

\textbf{Overviewing Multiple Targets} is similar to emphasizing a target. The key difference is that this purpose is aimed at observing multiple targets.
Although zoom is still a useful camera shot to present multiple targets as a whole, it sometimes causes overlaying issues with a huge amount of data and thus vision confusion. 
Therefore, video authors prefer to pan multiple targets individually at a finer granularity.



\textbf{Making a Comparison} uses another thing to explain or justify the main topic with similarities (\ie, analogy) or dissimilarities (\ie, contrast) in narratology~\cite{fludernik2009introduction}. 
It is also regarded as a common task for complex objects in data visualization~\cite{gleicher2011visual}.
Similar to geographic data videos, this narrative purpose only serves multiple geospatial targets.
Although overviewing multiple targets and making comparisons can be used to present multiple objects, we can differentiate them by the adjacent scenes.
A simple process of telling a geographic visualization is that designers first overview the map to introduce the context. 
Then, the map is zoomed in to emphasize an important region. 
Later, the map is zoomed out of the scene to compare similar data attributes between the highlighted part and the other regions.

\textbf{Supplementing Information} refers to circumstances that provide context information, such as adding additional descriptions and changing the viewing angle. 
Success in data visualization begins from building the context in the need of communication~\cite{knaflic2015storytelling}. 
In addition to presenting data insights, video creators sometimes need to add information that cannot be encoded in geographic visualizations to establish the context of the story. 
For example, if the author wants to deliver the information that Italy is a member of the European Union, they might add the icon of the European Union logo next to Italy on the map.
In most cases, such context information is inserted into the scene at the blank regions without information in the visualizations. Hence, a pan shot is the most used technique for this narrative purpose to save space for newly added information.

\textbf{Increasing Dynamics} refers to circumstances when no target is selected, and camera effects are used to give energy to the scene and create an atmosphere.
Satisfying this narrative purpose can be very simple but greatly increase the continuity of data videos. For example, zooming in/out the map subtly or just randomly moving the map.
Such scenes always happen at the beginning of the story or in a new section to introduce the relevant story and bring out the focus of the next story.

\subsubsection{Geospatial Targets}
The dimension of \textit{geospatial targets} describes \textit{what} type of visual object is commonly presented with the camera movements. 
Visual objects can be presented in single mode or in groups on the map.
We totally define six types of \textit{geospatial targets}. 
To describe the single object, we summarize three types of targets (\eg, location, region, and path) based on the geometric characterization of objects in geographic visualization.
In most scenes, including many geospatial targets, such design usually aims to explore the relationship among these geospatial targets regardless of their respective types.
Based on such findings, we group multiple targets (\ie, locations, regions, and paths) into a different type (\ie, multiple targets). Notably, the geospatial target can be \textbf{None} for a specific camera movement (\ie, \textit{Increasing Dynamics}).


\leavevmode\includegraphics[width=8pt]{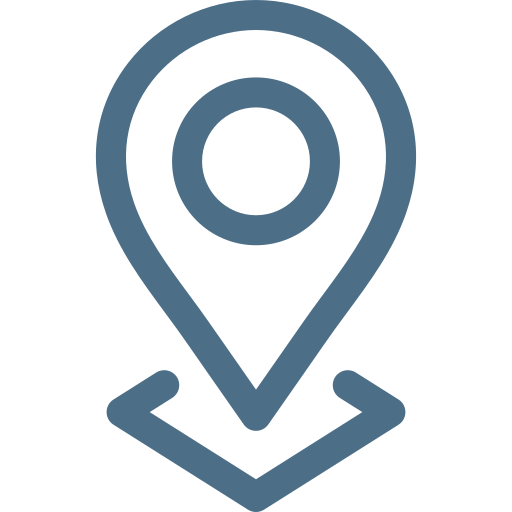}
\textbf{Location} has a point-like geometry respectively on the map. 
The type of target is the smallest one in the scene. 
Many camera shots can be used to present such a single location (\eg, a building or a bridge). 
For example, a zoom-in shot is a commonly used technique to change the narration topic from the last scene to the next single target.

\leavevmode\includegraphics[width=8pt]{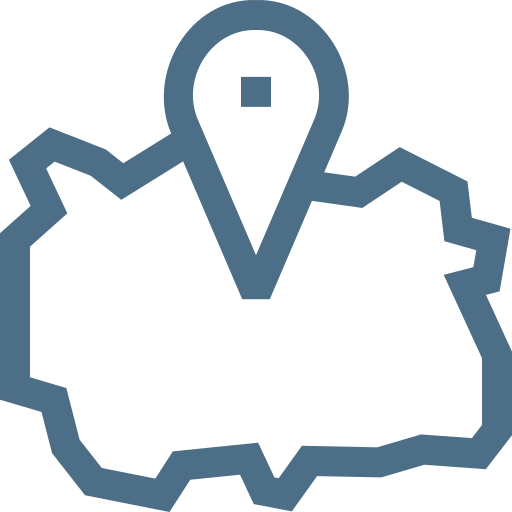}
\textbf{Region} has area-like geometry on the map (\eg, a country or a state). 
The region contains an additional area attribute than the location. 
Sometimes, the region can ignore its area attribute. Thus, region and location can be interchanged by visual appearance and camera movements. We can design similar camera movements for display. 
For example, if New York City is just marked to indicate where some events happened, then we can only zoom in on this visualization to emphasize its importance.
However, if we want to visualize the city's boundaries, then a pan shot, which is seldom used for presenting a single-location target, is a much more recommended technique. 

\leavevmode\includegraphics[width=8pt]{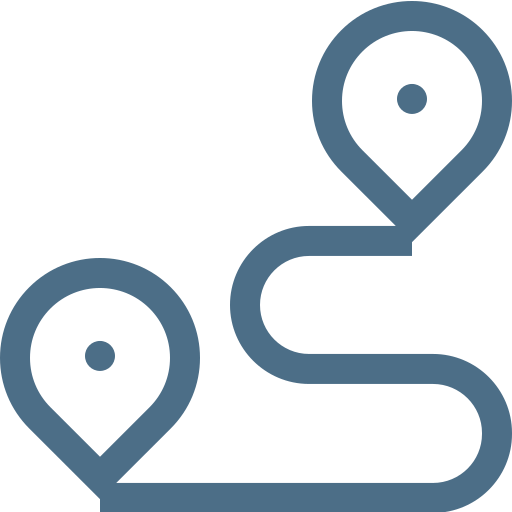}
\textbf{Path} has a line-like geometry on the map and is specific in the geographic domain (\eg, a river or the border). 
A path can show how connections among multiple locations. Thus, specific camera movements are used to have an overview and follow the cues of the whole path.


\leavevmode\includegraphics[width=8pt]{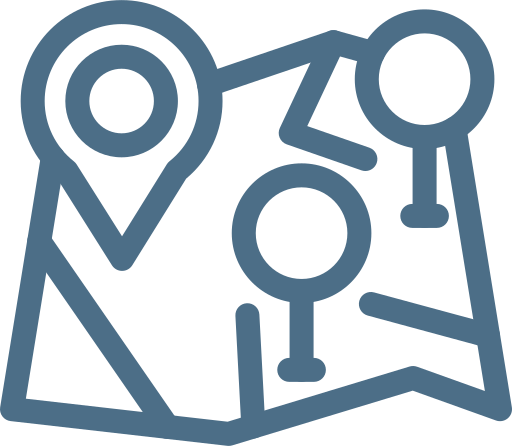}
\textbf{Multiple Targets} aims to describe a set of geospatial targets regardless of their respective types. 
When the designer presents many geospatial targets in a scene, they use the camera to construct the overview of or build a relationship (like comparison) between targets, which is related to the quantity rather than the type of targets. 
To be more specific, the scene should be designed to cover all the geospatial targets of this set.
For example, if a set of locations and another set of paths have a comparable geospatial distribution of visualizations (\eg, they are both within the borders of China), then it is highly likely to present two sets of targets on the map with similar scale and similar camera movements. 
Limited camera movements can serve multiple targets, such as panning among adjacent targets for comparison or zooming out to expand the scale of the map to overview all related targets.

\subsubsection{Camera Shots}
We summarized 9 types of camera movements regarding the camera's positioning, orientation, focal point, and moving path.
Some types are commonly used in real-world videos, and others are expanded from cinematography in films.
Our goal is to give inspiration for the design of camera movements.
Notably, one camera shot can serve many narrative purposes and geospatial targets with different parameter settings, such as the camera's moving speed.

\leavevmode\includegraphics[width=7pt]{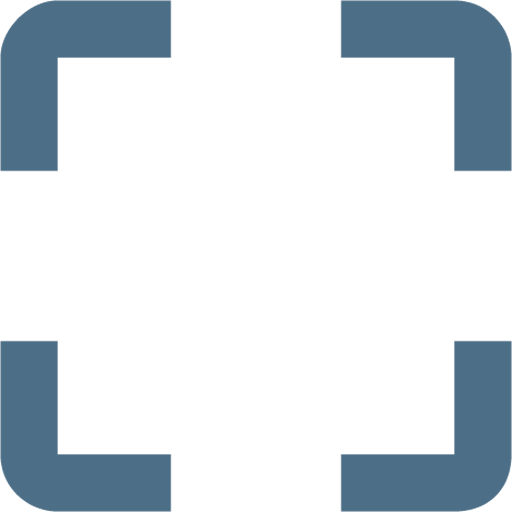}
\textbf{Static:} 
A static shot does not move the camera, hence resulting in an unchanged scene.
We still identified this type as a camera shot because this technique is frequently used in videos. 
The static scene can avoid audiences' confusion due to excessive many animations and direct their attention to the main content of the scene.

\leavevmode\includegraphics[width=7pt]{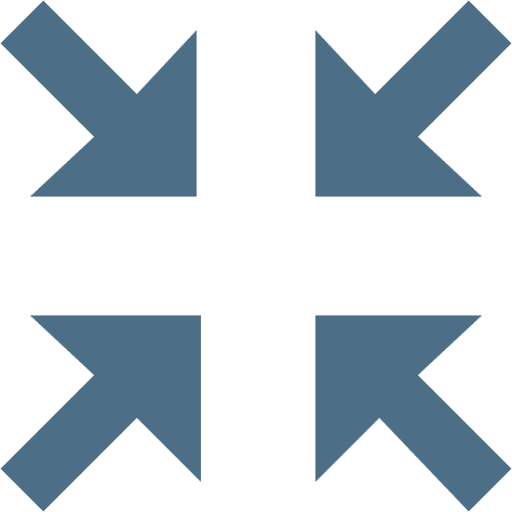}
\textbf{Push In:}
A push-in shot alters the positioning of the camera to become closer to an object. It is probably the most commonly used camera movement. 
It can reduce the map's scale to a specific geospatial target to draw the audience's attention toward the target. 
Especially, a push-in shot with a slow speed can increase the duration of displaying the target and leave audiences to expect what might happen ahead.

\leavevmode\includegraphics[width=6pt]{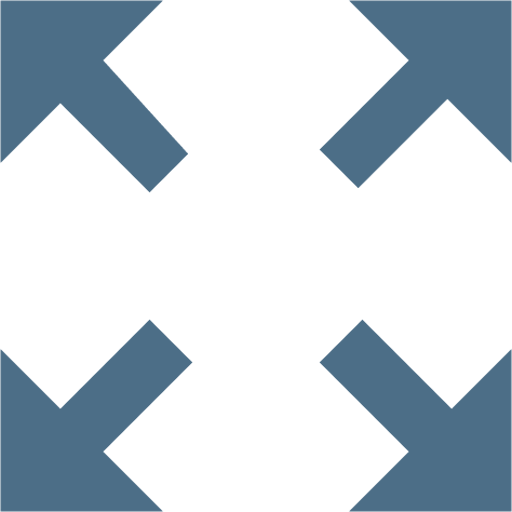}
\textbf{Pull Out:}
As opposed to push-in shots, a pull-out shot changes the camera itself to make it further away from an object. 
In this way, the map's scale enlarges to cover the adjacent objects and surrounding environment to provide a brief context of the target.

\leavevmode\includegraphics[width=7pt]{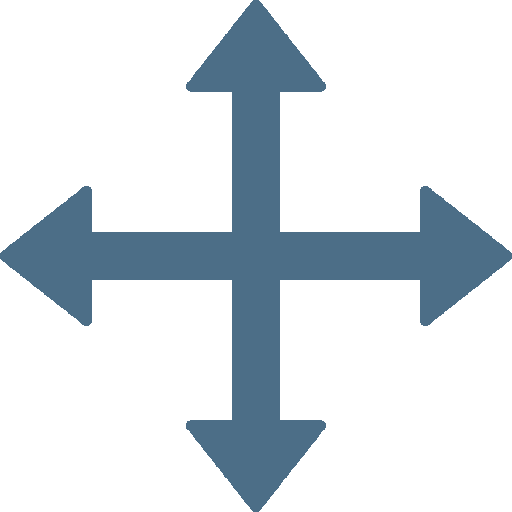}
\textbf{Pan:}
A pan shot refers to the camera that moves from one place to another.
In this way, the pan shot can change the focus of the map and thus change the targets in the scene.
Hence, it is usually used to change the topic of the story and emphasize the targets in the panned scene.
Panning multiple objects one by one can build an overview of all these objects.
Especially, this camera movement is identified as a whip pan when panning the camera with a quick speed. 
Different from panning at a normal speed, the whip pan ignores the information in the panning process. 
Therefore, it is handy for transitions that express the meaning of moving a large distance or time elapsing.

\leavevmode\includegraphics[width=7pt]{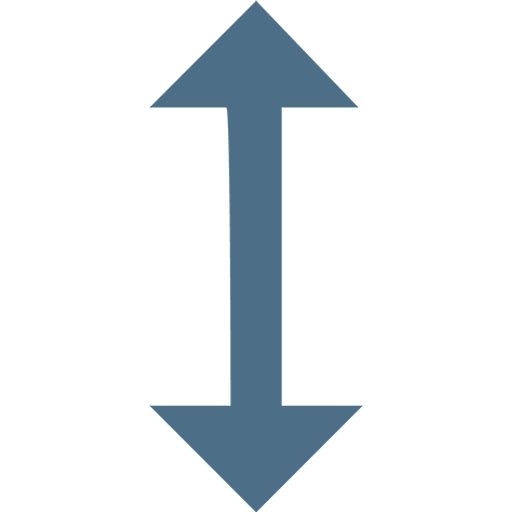}
\textbf{Tilt:}
Similar to pan movements, a tilt shot moves the camera shot vertically upward or downward. 
The tilt shot, as an unveiling technique, is helpful either to reveal from top to bottom or the reverse. 
For example, it can be used to describe the bar height from its bottom to the top to emphasize the data values encoded by the bar.

\leavevmode\includegraphics[width=7pt]{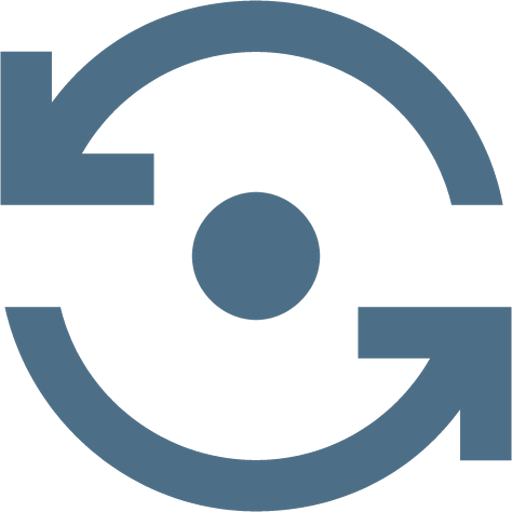}
\textbf{Camera Roll:}
A camera roll rotates the camera pointed at the same object on its long axis.
The rolling camera can emphasize the object from different perspectives.

\leavevmode\includegraphics[width=8pt]{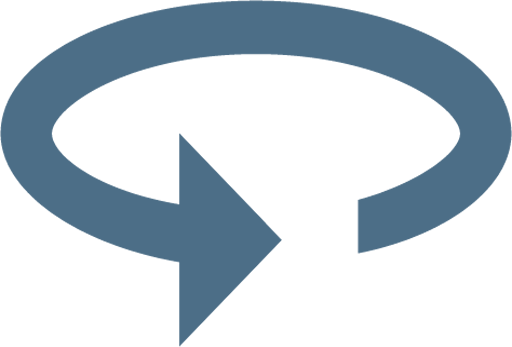}
\textbf{Arc:}
An arc shot moves the camera around the same object in an arcing orbit.
It is typically used to add dynamics to a static object for emphasis because of its longer duration.

\leavevmode\includegraphics[width=7pt]{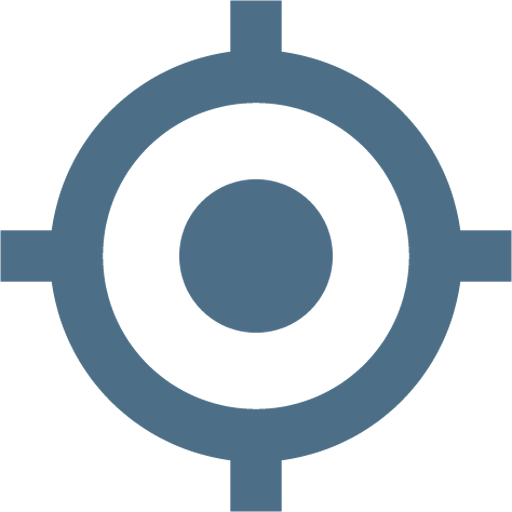}
\textbf{Tracking:}
A tracking shot describes any shot that moves alongside a subject for a period of time.
It can be used to simply follow a path and thus display detailed information, such as nearby cities and the surrounding transportation. 


\section{\tool Overview}
\label{sec:system}

\subsection{Design Considerations}
\label{sec:design-consideration}

The authoring tool is designed to simplify the process of crafting geographic data videos for the general users. 
We summarize a set of high-level design considerations for an authoring tool that empowers average users to craft the camera movements in geographic data videos quickly based on literature survey and design space study.
We assume that the video makers have attained sufficient insights into geographic data, and their task is to organize these insights into a coherent story.

\noindent\textbf{(C1) Facilitate easy camera authoring with narrative purposes.}
Average users may not be familiar enough with diverse camera shots to create a compelling geographic data video.
To assist them in the authoring processes, narrative purposes, such as emphasis and comparison, should be defined to empower users to design camera movements at the semantic level. 
\rev{For example, when a user try to compare two regions, they simply select the comparison purpose and zoom into the regions they want to compare.} 
Narrative purposes abstract away the details in configuring camera shots and offer users an easy and intuitive way to convey geographic insights with cameras. 


\noindent\textbf{(C2) Suggest appropriate camera parameters adaptively.}
Different from general data videos, videos in geographic visualization are more concerned about the performance stage (\ie, the proportion of objects in the scene and the viewing perspective~\cite{shi2021animation}).
Suppose the designer wants to show the highest bar in a bar visualization on the map.
In that case, the scene will contain the whole bar visualization as the indispensable context (\eg, ~\cite{geovideo2}).
Non-experts easily get confused about editing obscure parameters of camera movements to choose appropriate performance stages for the selected geospatial targets.
We should recommend adaptive graphics parameters for camera movements on the basis of geospatial targets themselves and the related geographic environments on the map.

\noindent\textbf{(C3) Depict the narrative timeline of a geographic data video.}
In geographic data videos, designers always focus on exploring the spatial relationship and presenting spatial context. 
General video authoring tools (\eg, Adobe After Effects~\cite{aftereffect}) typically support keyframe-based specifications for authoring the animation.
However, traditional keyframes cannot guide users about the spatial context in geographic visualizations.
We aim to preserve every spatial context users tend to narrate to highlight geographic visualizations.
We propose a hierarchical timeline, including time, spatial context, and camera movements.
Such a timeline allows users to recognize the current narration sequence and edit the camera and its duration based on the spatial context.



\subsection{Video Modeling and System Workflow}

\begin{figure*}[th]
    \centering
    \includegraphics[width=\textwidth]{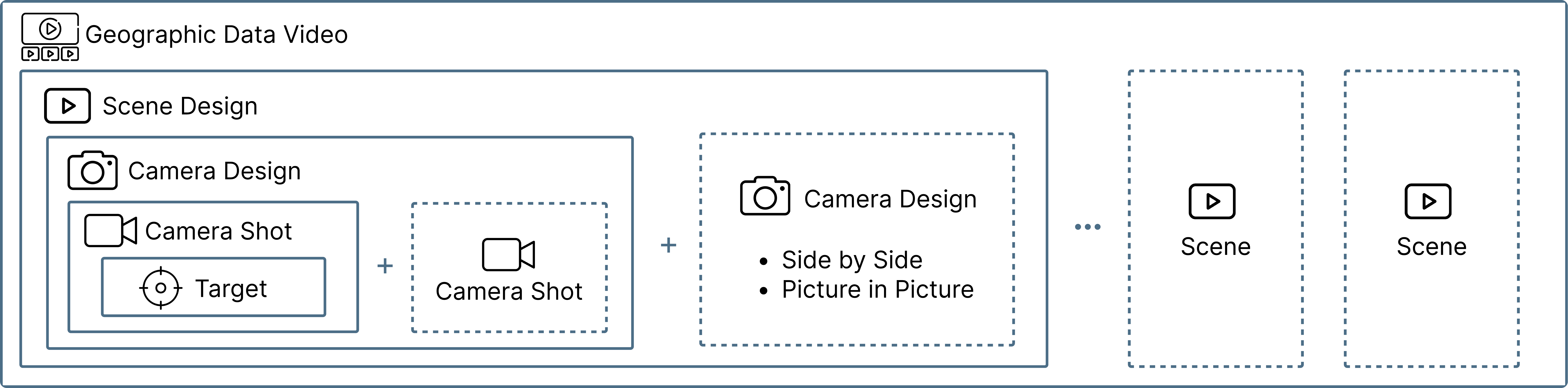}
    \caption{The video model in GeoCamera. A geographic data video is a series of scenes. Each scene comprises one or multiple camera designs. Each camera design includes one or two camera shots. Each camera shot serves a geospatial target.}
    \Description{Figure 3. Fully described in the text. }
    \label{fig:model}
\end{figure*}

As illustrated in \autoref{fig:model}, a geographic data video can be defined as a series of scenes. Each scene comprises one or more camera designs combined by using layouts, such as side-by-side and picture-in-picture. For each camera design, a camera shot or a combination of camera shots is chosen to present one or multiple geospatial target(s) with one of the narrative purposes. 

Given the geographic visualization, \tool first needs to generate a list of camera movements used for specific geospatial targets. 
Users first select the geospatial targets to be presented in the scene and then choose a set of camera movements for each target.
\tool records the selections and visualizes them into a location-camera hierarchy timeline.
\tool supports flexible interactions to edit semi-automatically generated concrete camera effects individually in the data video from the graphics parameters and the timeline.
Lastly, \tool assembles the list of camera movements into a data video.

\subsection{Creating Camera Movements}
\subsubsection{The main previewer of interactive geographic visualizations.}
\tool provides a canvas (\autoref{fig:ui}A) to visualize the geographic visualization and preview camera movements in the map individually and sequentially. 
In the beginning, this canvas only contains an interactive map without any encoded data as an initial hint of geographic visualizations.
After users import their geographic data and select a geographic visualization type (\eg, hexagon), the tool will draw the selected visualization on the map with predefined visual attributes (\eg, color).
Users are allowed to adjust the perspective and scale of the map (\textit{Basic Control}) and select their preferred geospatial targets (\textit{Interactive Selection}) through basic controls and target selections.

\textbf{Basic control.}
Consistent with most map authoring tools (\eg, deck.gl~\cite{deck}), we use an interactive map through the whole authoring process.
Users can zoom the map by scrolling.
They can also pan and rotate the map by clicking and dragging the cursor.
These primary interaction handlers allow users to adjust the map to a favorite state in a non-programming way. 

\textbf{Target selection.}
With basic controls, users can opt for a single object by clicking it on the interactive visualization. The location of the object will be recorded based on the underlying geographic map. The user can also select a specific region or multiple targets by using a lasso selection (\autoref{fig:ui}A). Besides, we found users would probably alter the map mistakenly with imperceptible mouse operations, thereby making it challenging for the users to select targets under the same viewpoints. Therefore, we designed an error-tolerance mechanism to ensure that the viewpoints stays in the same state as it was before. 
Users can save the current viewpoint by clicking the ``snapshot'' button and then go back to the previously saved map state by clicking the ``reset'' button (\autoref{fig:ui}A). 

\subsubsection{Library of camera shots.}
\begin{figure}[ht]
    \centering
    \includegraphics[width=\columnwidth]{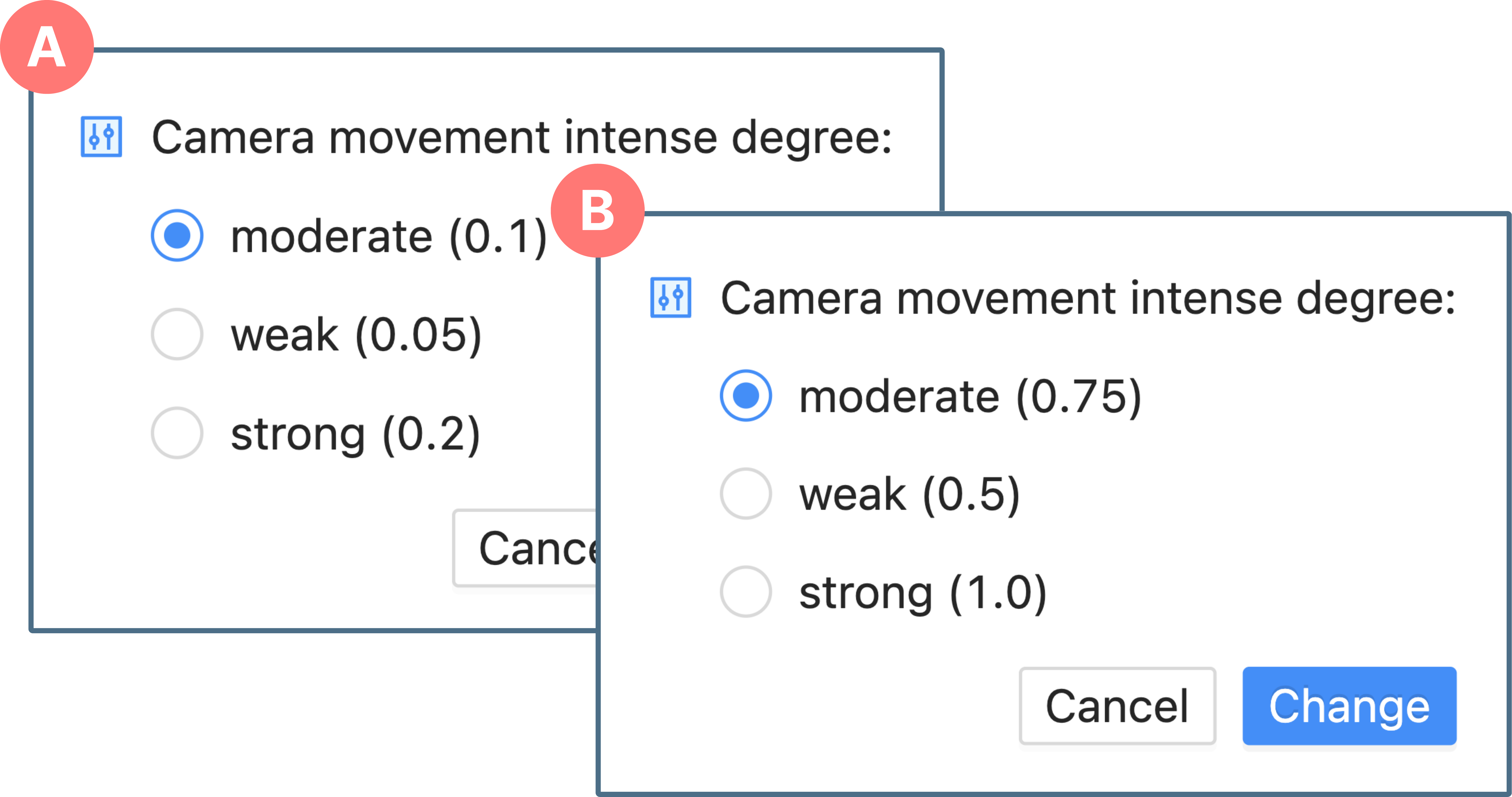}
    \caption{Different suggested options for a \emph{Push In} shot that serve different narrative purposes: (A) \textit{Emphasizing a Target} and (B) \textit{Increasing the Dynamic}. Different intensities of the movement influence the final effects of the camera movement.}
    \Description{Figure 4. Fully described in the text. }
    \label{fig:narrative}
\end{figure}

Based on the previous investigation of real-world videos, we designed a library of camera shot templates (\autoref{fig:ui}B).
To facilitate the camera effect authoring for users, we create different catalogs for various narrative purposes defined in the design space. 
Each catalog includes several camera shots collected and summarized from the examples.
Note that different camera movements that belong to the same type can serve different narration purposes with various parameters.
Additionally, we expand the current categories inspired by films~\cite{katz1991film, vineyard2000setting} to enrich camera shots, such as combined camera movements (\eg, \textit{Arc} with \textit{Tilt} shot). 
For the same camera shot serving different narrative purposes, the system provides different configuration settings for users (\autoref{fig:narrative}) to achieve their narrative goals.

Based on the observation from the example videos that a central object can be served by different camera movements, \tool enables users to craft camera movements one by one for specific geospatial targets. 
After users determine their targeted objects, they can select a camera shot to describe these targets for storytelling. 

For example, if a user wants to build an overview of multiple objects in the geographic visualization, then they need to select a camera shot under the narrative purposes of ``Overviewing multiple targets'' in the library. 
A default camera shot will be added to the timeline after the user selects the target in the geographic visualization and the narrative purpose. The default camera shot is determined on the basis of the frequency statistics of the collected real-world examples. \rev{As shown in \autoref{fig:corpus}A}, the most frequent camera shots will be chosen as default when the narrative purpose and geospatial target is confirmed. The default camera movement for the targets will be automatically created on the timeline (\autoref{fig:ui}D) after a few clicks and will be documented in the camera list (\autoref{fig:ui}C). The user can change the camera shot type for the camera movements by clicking the buttons on the camera shot list. By repeating this process, users can build their whole narration with a sequence of camera movements using their corresponding geospatial targets. 

\subsubsection{Adaptive parameter setting for camera movements.}

The camera movement in our system depends on the narrative purpose, the camera shot, and the geospatial target selected. It is achieved by interpolating from the initial state to the final state of the camera. However, configuring the states of a camera is a challenging and time-consuming task for a lay person. Thus, we introduce an automation service that adaptively sets up camera movement parameters for the chosen geospatial target(s) after the user determines the camera shots under a specific narrative purpose.
This step waives the user's setting of a camera's states, which largely simplifies the camera movement creation with only a few clicks. 

The initial camera state of a camera movement can be set by considering the current viewpoints or the final camera state of the last movement. Meanwhile, the final state of the camera is dynamically adjusted by the three aspects of the selected geospatial target(s): the centroid, the bounding box, and the related data of the selection. 
The centroid is generally used to define the final location and the focus point of a camera. Moreover, the bounding box and the data of the selected target(s) decide the altitude of the camera. 

If the user selects a location target, then the centroid is its geographic coordinates. For the region target case that is selected by a lasso selection tool in the UI, we calculate the centroid of the polygon. Especially, for a path target or multiple targets, we compute the bounding box first and then obtain the centroid based on the bounding box. 

After deciding the location and focal point of a camera, we need to achieve an appropriate viewport for the target. The core idea is to ensure that the viewport covers the target(s) the user wants to observe. For instance, for a push-in shot, the final state of the camera movement will be set to vertically focus on the target or the centroid of the bounding box. At the same time, the altitude of the camera after the push-in movement will be guaranteed to a value that satisfies a margin space for each side at 10\% or above. Similarly, we heuristically define a set of rules for the eight camera shots to adjust the viewport of the camera dynamically based on different geospatial targets. 


\subsubsection{Location-camera hierarchy timeline.}
In addition to playing a single camera movement, \tool also supports playing all camera movements sequentially.
In the timeline panel (\autoref{fig:ui}D), we visualize two timelines: a general timeline and a location-camera hierarchy timeline.
In general, we use a simple slider to show the playing process of all cameras in the camera list.
After users click the triangle button above the general timeline, the previewer will play the camera on the geographic visualization, and the general timeline will visualize the current playing position.

However, the general timeline only shows the temporal information of the entire camera sequence.
We design a location-camera hierarchy timeline to ensure the temporal information of each camera movement and each geospatial target. 
We draw a timeline for every camera in the camera list, including a fixed range encoded for the duration of the camera.
To help users obtain geospatial targets for each camera movement, we aggregate cameras with the same geospatial targets and add a location-level timeline to show how long the video focuses on the same targets.
All the location-level timelines are arranged chronologically. Therefore, camera-level timelines are in the same order. 
Users can drag marks on camera-level timelines to change the cameras' duration.
When creating a story, having a break before starting a new topic is common, but continuous sentences are used within a topic.
Based on such considerations from the narrative aspect, we identify a rule for continuity between two continuous timelines when editing duration.
An interval without cameras between two continuous location-level timelines is possible.
However, the interval between two continuous camera-level timelines under the same location layer is restricted. 
We also design a semantic overview of all timelines.
Location-level timelines are named with their related locations.
If we cannot obtain the exact name from the import data, the tool will show the longitude and latitude of the location.
Camera-level timelines are named as their corresponding camera movements with their narrative tactics.

\tool smooths transitions with linear interpolation among camera movements.
When users craft the camera movements, they adjust the duration of movements at a single level.
In terms of the final data videos, we set the rule to avoid the temporal contradiction of camera movements for different locations.
However, an interval can appear between the ending state of the last camera movement and the initial state of the next movement.
Considering the videos' coherence, we provide a special camera movement called ``linear interpolation'' to fill in the gaps in time.
Different from other camera movements, this movement focuses on no geospatial targets.
Its performance in the scene is the same as ``fly to,'' an automatic map animation that is commonly used in map visualization tools (\eg, Mapbox~\cite{Mapbox}).

\subsection{Configure Visual Effects}

\begin{figure}[t]
    \centering
    \includegraphics[width=\columnwidth]{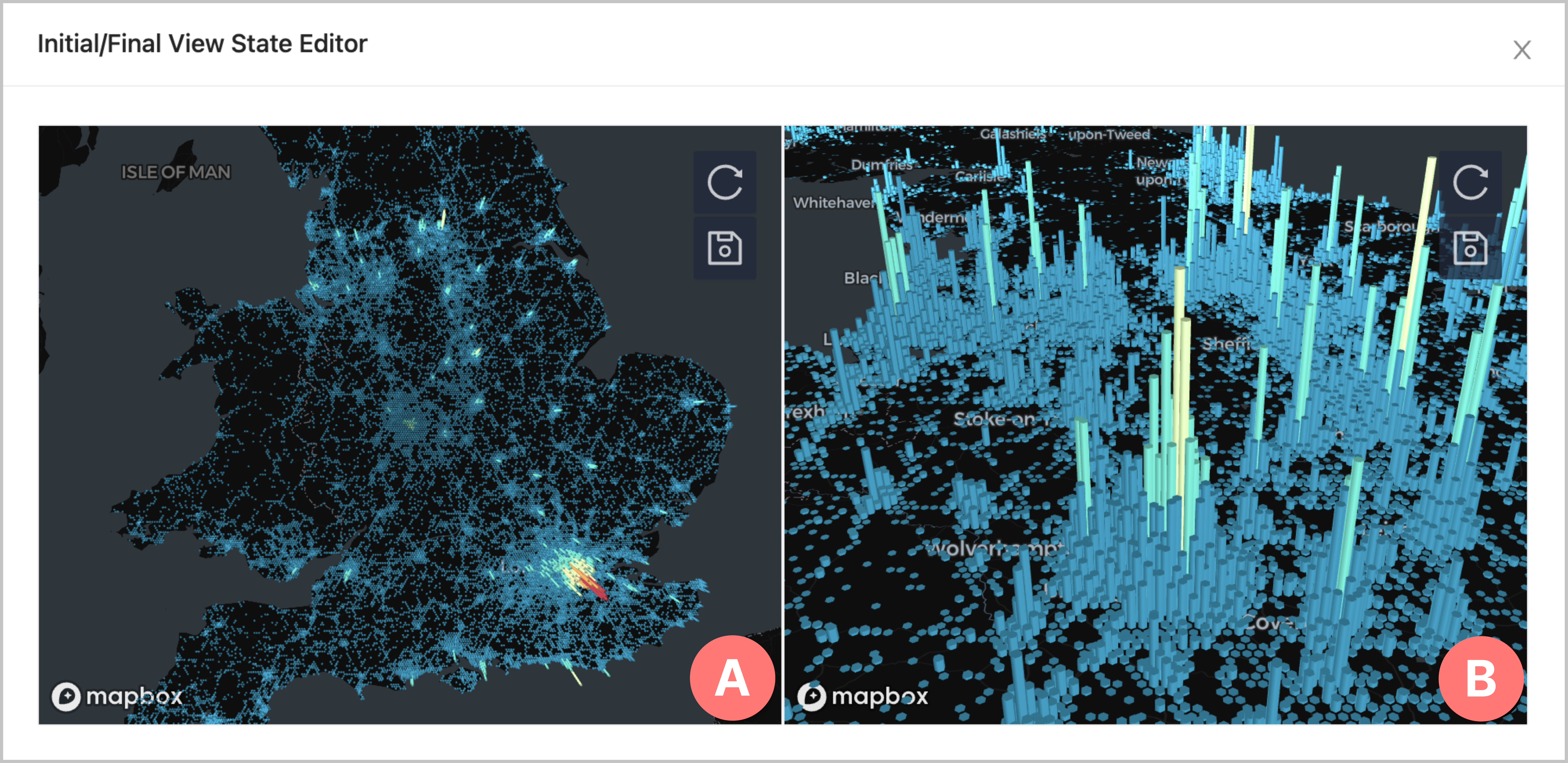}
    \caption{The user can edit a \textit{Tilt} shot with two single previewers: (A) one is used for editing the initial state of the camera movement, and (B) another is for the final state. Then, the camera movement is interpolated between the two states.}
    \Description{Figure 5 shows the editing interface for the initial and the final states of a camera movement. }
    \label{fig:camera-edit}
\end{figure}

\subsubsection{Camera configuration with single previewers.}
Although \tool{} recommended adaptive camera movements, users may be not satisfied with the results.
Based on the demand for flexible refinement, \tool{} enable users to edit every camera movement's graphic parameters in the camera list (\autoref{fig:ui}C).
At the start, all camera parameters are folded up. 
If users want to edit a camera movement, then they will click its menu to expand its parameterization panel.
These parameters are professionally related to the map, such as zoom, pitch, and bearing. 
Users who are not familiar with the geographic domain can easily get stuck in understanding these parameters.
Therefore, \tool{} provides a single previewer approach to help users skip this understanding process of refining a camera movement.
By clicking the ``edit state'' button, \tool displays a small view with two canvas (\autoref{fig:camera-edit}). 
One canvas shows the state of the geographic visualization in the previewer when the camera starts moving, and another is for the ending. 
Both canvases have similar basic interaction handlers as the main previewer. 
Users adjust the states of two canvases with the mouse until they are satisfied.
\tool records the adjusted states and resets the parameters of the camera movement based on the recording.

To guide users to have a preliminary concept of how the camera movement works, \tool{} allows them to play each camera movement individually.
If users want to watch a camera movement while not completing a story, they just need to click the small triangle button in the parameterization panel of this camera.
Then, in the previewer, \tool plays the change in the geographic visualization using the camera movement from the initial state to the ending state.

\subsubsection{Visualization authoring.}
Our tool aims to provide an easier way to craft the camera animations in the presentation of geographic visualizations.
This method requires an accomplished geographic visualization as the input.
However, our target audiences include those non-experts in the geographic data analysis domain. 
They are unfamiliar with selecting the proper visualization types and drawing the selected visualization on the map.
With this in mind, \tool{} wraps graphing functions to support the authoring of several common visualizations from the cleaned geographic data.
Corresponding to the summarized geospatial targets before, these visualizations include point, three-dimensional rectangular, and line map.
After importing geographic data (\ie, every object has its exact latitude and longitude), users select an appropriate geographic visualization according to data type.
Users change the input of visualization type with the drop-down menu in the Configuration Panel (\autoref{fig:ui}C), and then \tool{} displays the selected visualization in the Preview Panel (\autoref{fig:ui}A).
In this way, users only need to import the geographic data rather than a geographic visualization, which reduces the threshold of using \tool{}.

The output of \tool{} is a geographic data video that assembles all camera movements.
However, a successful data video contains visual and auditory stimuli~\cite{amini2017dataclips}. 
Camera animation is a small component of visual design.
Although our work focuses on camera movements in geographic storytelling, \tool{} provides an additional layer for annotations for design enhancement. 
Users can click the ``Annotation'' button in the Configuration Panel (\autoref{fig:ui}C), and the panel will display all the current textual annotations that correspond to each camera movement aligned by the time sequence.
The default annotation of a camera movement is none. 
The User can choose a camera movement and edit the text. The tool will add an annotation layer for this camera. 
We align the time of the added annotation layer with its camera.




\begin{figure}[t]
    \centering
    \includegraphics[width=\columnwidth]{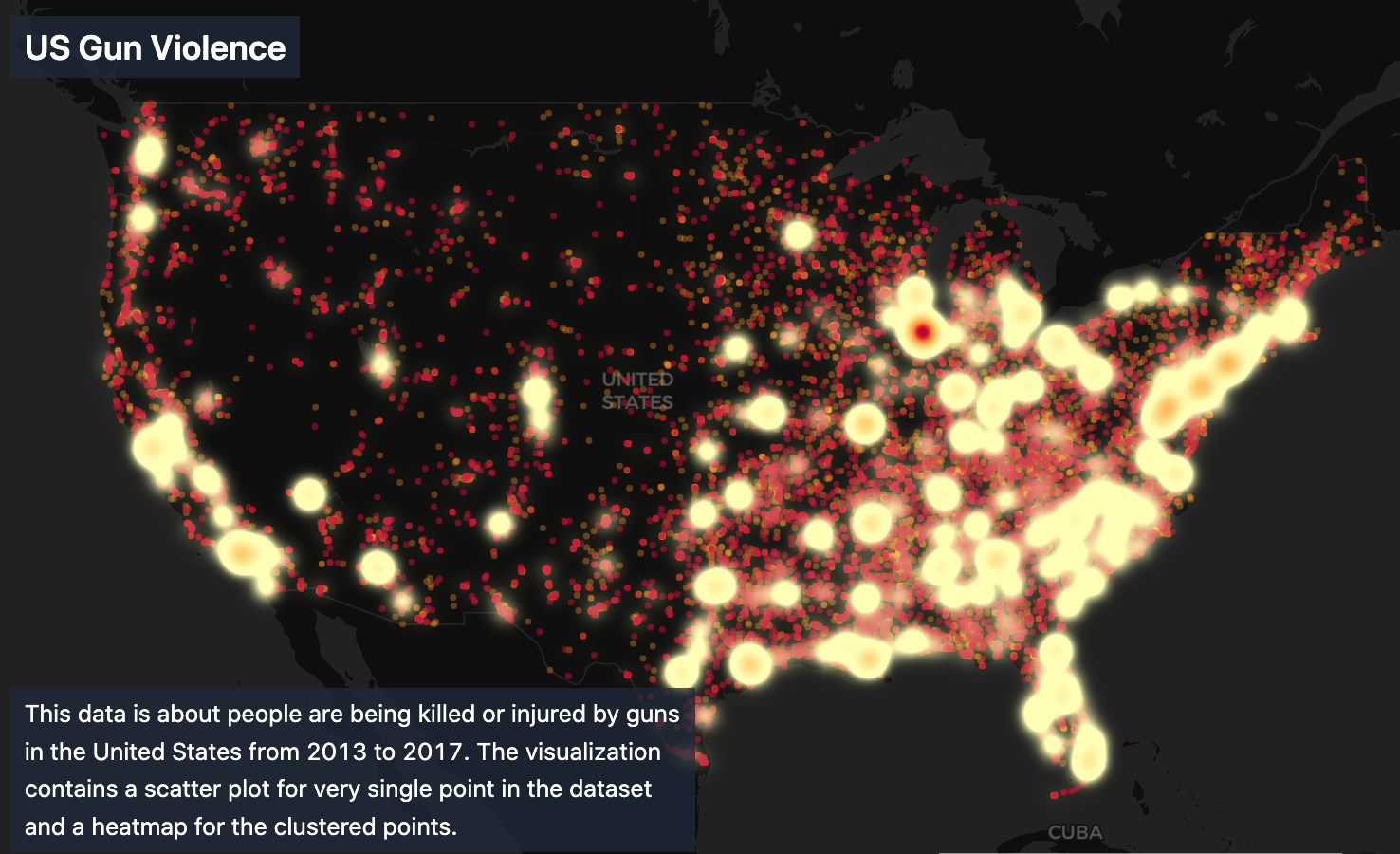}
    \caption{The data visualization about gun violence from 2013 to 2017 in the United States.}
    \Description{Figure 6 shows a snapshot of the data visualization about gun violence. }
    \label{fig:us-gun}
\end{figure}

\section{Evaluation}
\label{sec:evaluation}
We evaluate \tool{} through (1) two example cases with a range of camera shots to showcase its expressiveness and (2) a user study to verify its usability.

\subsection{Example Cases}
To demonstrate the generalizability of \tool{} in different geographic visualizations, we select a 2D and a 3D geographic data videos as our use cases. The geographic visualization in the first example is a combination of heatmap and scatter plot, showing more than 140,000 data points of people being killed or injured in America. This visualization shows location-based and region-based geographic data story. The data video is created based on the geographic data story.  Another example is a visualization showing COVID-19 cases in Hong Kong. We replicate a designer-crafted real-world data video, exaplaining the case distribution. The video contains railway network map and scatter plots on map and utilizes camera movements (\eg, \textit{Tilt} and \textit{Arc} shots) for diverse geospatial targets of locations, paths, and regions that are rarely observed in the 2D environment, thereby complementing the types of camera movements in the previous example. 


\subsubsection{US Gun Violence}
In this case, we attempt to use \tool{} to create a data video to present insight from the data about people killed or injured by guns in the United States from 2013 to 2017. The visualization contains a scatter plot for every single point in the dataset and a heatmap for the clustered points (\autoref{fig:us-gun}). At the beginning of the video, we provide an overall introduction to the dataset. We apply the frequently used \emph{Push in} shot in the \textit{Increasing Dynamics} category (\autoref{fig:corpus}) to gradually immerse the audience in the visualization scene. The camera movements for this category are mild, thereby suggesting a long duration \emph{Push in} for a short-distance. The camera movement is set to last 10 seconds with the annotation showing \says{The data are about recorded gun violence incidents in the US between 2014 and 2017.} Subsequently, we observe that the eastern United States has a relatively higher chance of shooting than the other areas of America. We select the eastern part to present this observation. We again use a \emph{Push in} shot, but in a different category of \textit{Overviewing Multiple Targets}. This time, the camera pushes in relatively quickly and focuses on the selected region of America within 2 seconds, and then remains for another 5 seconds to show the annotation \says{The visualization indicates that eastern America has a higher chance of shooting incidents than the other parts of America.} Then, we compare the overall shooting rate between eastern America and middle America. We select the middle part in the visualization, and the system automatically suggest a \emph{Push out} shot in the timeline after we select the \textit{Making a Comparison} category to show the two regions together at the same time. The camera movement takes about 2 seconds, and remains for another 2 seconds with the annotation showing \says{However, the middle part of America is relatively safe.} Subsequently, we aim to show another observation that the number of gun-related violence outstands in California. We replace the default camera shot (\ie, \emph{Push In} shot) with the \emph{Camera Roll} in the \textit{Emphasizing a Target} category to roll the camera while maintaining its focus on the center of California. This eight-second camera movement is about \says{The number of gun-related violence evidently outstands in California,} and the audience is provided sufficient time to understand the surroundings of the mentioned State. What follows next is the most dangerous city with the highest number of shootings. We further use another \emph{Push In} shots in the same category to draw attention to Chicago. The camera gradually moves to Chicago for an additional 3 seconds to show that \says{Unexpectedly, Chicago is the most dangerous city with the highest number of gun-related violence.} Finally, we select a \emph{Pull out} shot in the \textit{Increasing Dynamics} category to leave the visualization scene in 10 seconds, \says{Overall, there was nearly 226 thousand gun-related cases were recorded in the US, and 60 thousand people were killed.} 
The total duration time of the final video with the six camera movements is 43 seconds. The time distribution of the corresponding six narrative purposes (\ie, \textit{Increasing Dynamics}, \textit{Overviewing Multiple Targets}, \textit{Making a Comparison}, \textit{Emphasizing a Target}, \textit{Emphasizing a Target}, and \textit{Increasing Dynamics}) is shown in the timeline, indicating the story flow of the generated video. More details are presented in the supplementary material. 

\subsubsection{Reproducing Camera Movements}
\begin{figure}[t]
  \centering
  \includegraphics[width=\columnwidth]{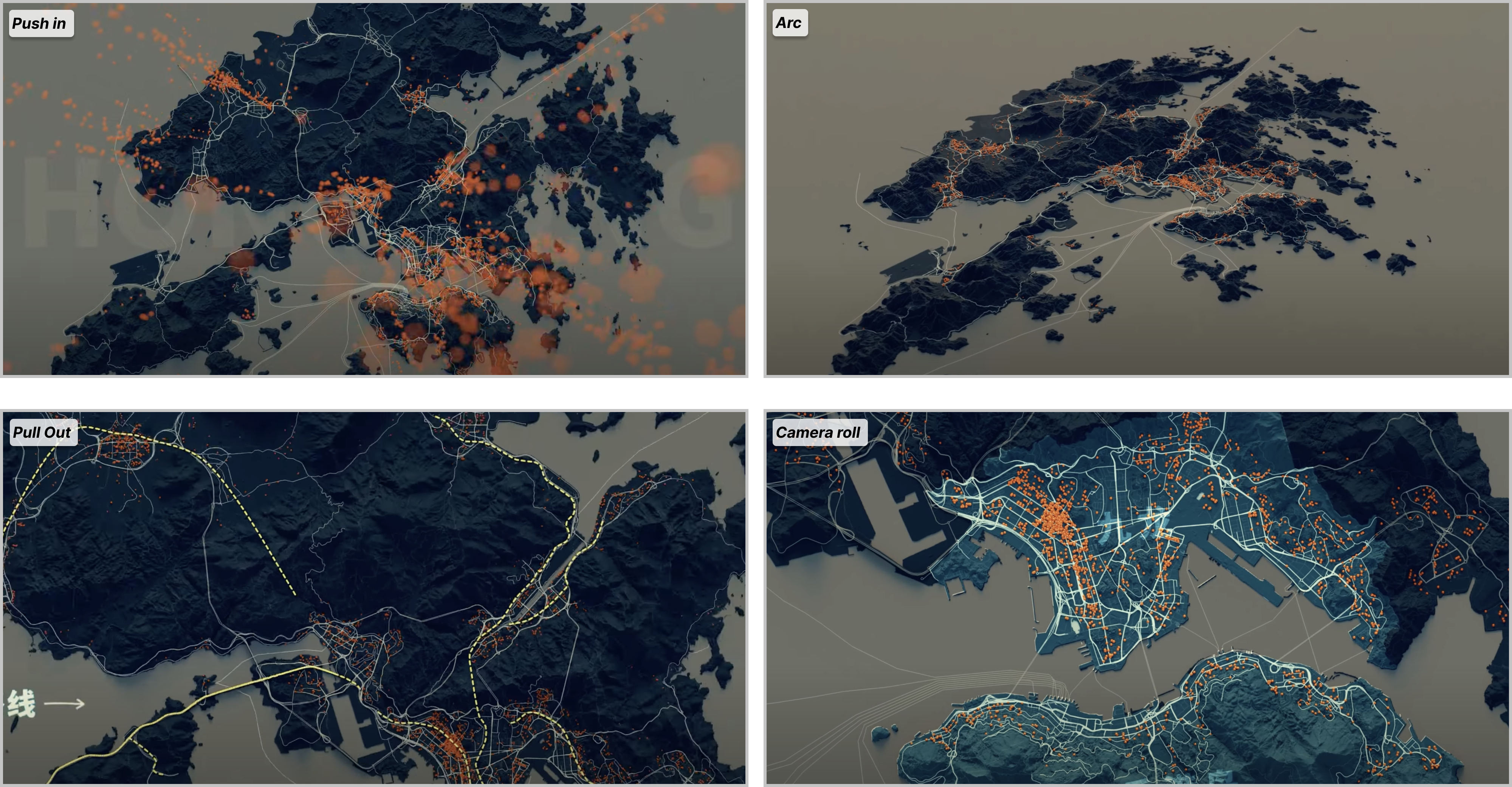}
  \caption{The snapshots of the sample camera movements taken from the real-world geographic data video. The camera shots identified in these camera movements: \textit{Push In} shot, \textit{Arc} shot, \textit{Pull Out} shot, and \textit{Camera Roll} shot.}
  \Description{Figure 7. Fully described in the text. }
  \label{fig:reproduction}
\end{figure}

To demonstrate the expressiveness of the camera movements in our system, we reproduce the camera movements in a data video introducing the COVID-19 status in Hong Kong during February 2022. 
We use the manual mode that does not require specifying the narrative purposes for the camera movements, given the circumstance that the camera effects have already been provided by the sample video. The manual mode is designed for the detailed control of generating specific camera effects from scratch without considering the narrative purposes. 
The data video contains 32 camera movements covering the \emph{Push in}, \emph{Arc}, \emph{Camera roll}, \emph{Zoom out}, and \emph{Pan} shots (\autoref{fig:reproduction}). The set of camera movements form a three-minute data video. More details about the reproduction results are shown in the supplementary material. 

\subsection{User Study}
We conducted a user study to validate whether or not nonprofessionals could easily create different camera movements for a data story and examine if there were any usability issues for improving the system. 

\subsubsection{Participants}
We recruited eight participants (two females and six males) with knowledge in geographic visualization, denoted as P1-P8, for this study. They involve graduate students studying data visualization and employees that work at an IT company in a role related to data visualization. All the participants took no part in any activity related to the system design or preliminary study and reported that they had no or limited experience in camera movement design. 

\subsubsection{Visual Materials and Data}
We provided the participants with slides that introduced each category of our design space with examples. The slides were used as the teaching material, and the participants were encouraged to browse the slides when creating camera movements with our system. 

The data used to create data videos in the user study include personal injury road accidents in the United Kingdom from 1979. The data are aggregated and visualized with a hexagon-based heatmap (\autoref{fig:ui}(a)). The color and height of a hexagon are determined based on the objects it contains. In addition to data visualization, we also provided the participants various pre-extracted story insights (\eg, \textit{``London has the most road accidents''} and \textit{``The road accident in Scotland area is very low''}) and their corresponding visualization to maintain the focus of the user study on planning and creating camera movements. 

\subsubsection{Procedure}
The user study contains three sessions, as follows: (1) a tutorial session to familiarize with \tool{}, (2) a creation session to experience the authoring process, and (3) a post-study evaluation to measure their subjective preference on the utility of the system.

\textbf{Tutorial.} 
We started the user study with a 15-minute introduction explaining our design space. Subsequently, we provide a 20-minute demonstration on the use of the \tool{} system, including its functions for setting up camera movements and related interactions in camera movement editing with an example dataset. 
Then, the participants were asked to freely explore every function as well as the interaction of the system and raise questions whenever necessary. 
After familiarizing with the tool, we introduced the formal dataset for the user study, the encoding scheme of the proposed visualization, and the corresponding story insights in detail for building camera movements with \tool{}. 

\textbf{Creation.}
After the tutorial, we asked the participants to use \tool{} to make their own geographic data videos based on the pre-extracted story insights. The participants could browse the slides as a reference, and they can seek guidance for using the system. When finished, each participant shares and explains his or her data video. The creation phase lasts for approximately 15-30 minutes. 

\textbf{Post-study Survey and Interview.}
When the exploration and creation of the camera movements were completed, the participants were asked to answer a post-study questionnaire using a 5-point Likert scale (1 for strongly disagree and 5 for strongly agree). The questionnaire intends to assess the usefulness, ease of use, and satisfaction\cite{lund2001measuring} of \tool{}. Finally, we conducted a semi-structured interview to collect qualitative feedback from each participant. 

All participants completed the entire study at approximately 75-90 minutes and were compensated with a gift card worth \$15 at the end of the interview session. 

\begin{figure*}[th]
    \centering
    \includegraphics[width=\textwidth]{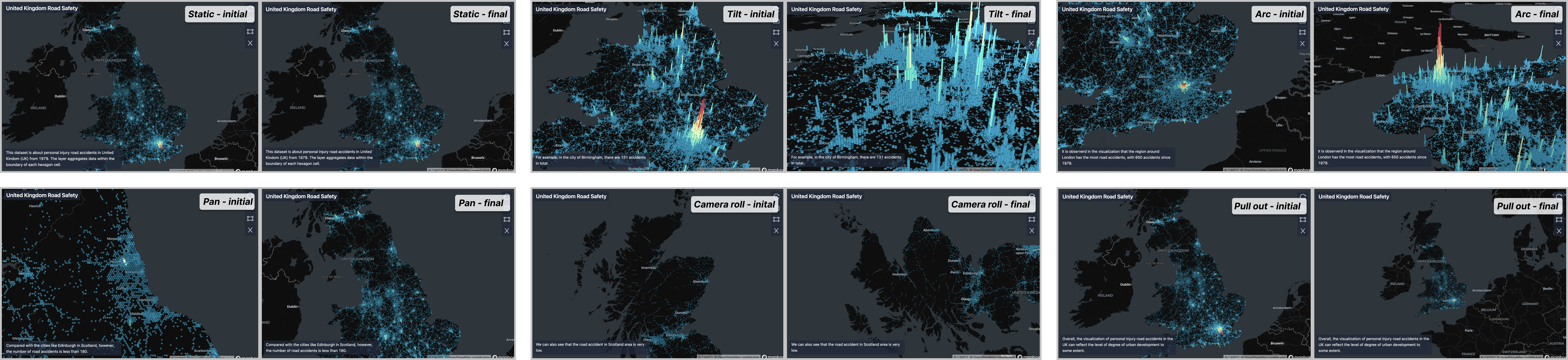}
    \caption{Video snapshots from the user study generated with GeoCamera. Each camera movement contains two snapshots of the initial and the final state. }
    \Description{Figure 8 shows the snapshots of a user study example with six camera movements generated with GeoCamera. }
    \label{fig:study_example}
\end{figure*}

\subsubsection{Results and Findings.} 

\begin{figure*}[ht]
    \centering
    \includegraphics[width=\textwidth]{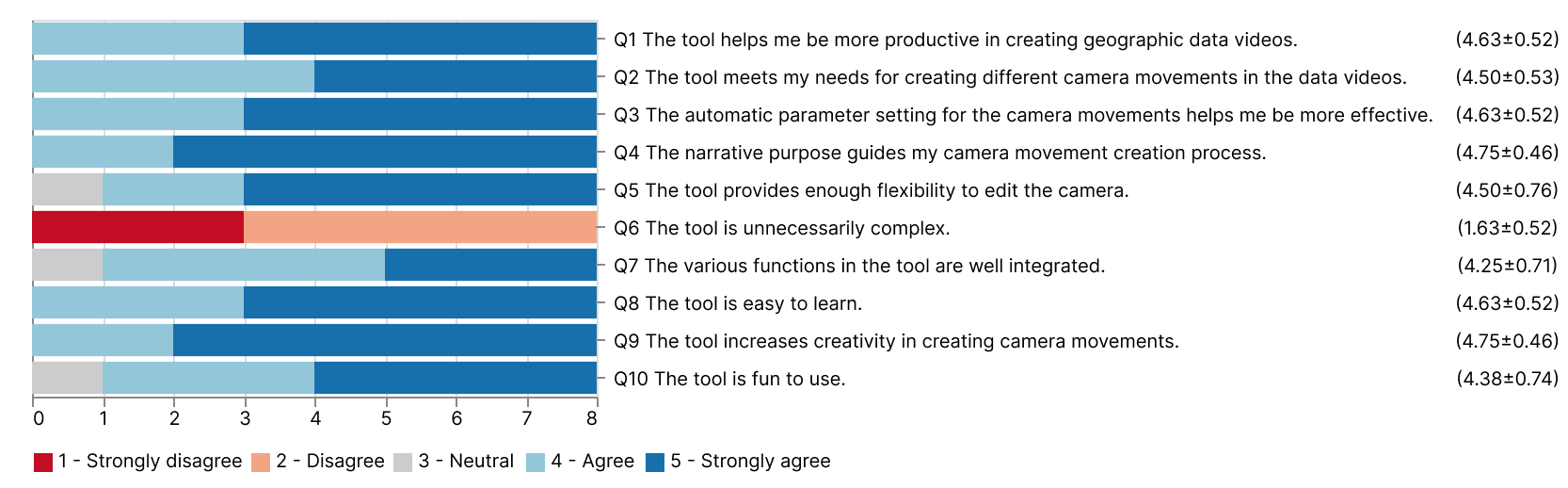}
    \caption{Ratings for system usability on a 5-point Likert scale (N=8). The middle column shows the detailed questions. The right column displays the average and standard deviations. }
    \Description{Figure 9 shows the questions for system usability and their rating distribution from the user study. }
    \label{fig:likert_result}
\end{figure*}

All the participants can complete the authoring tasks with minimal guidance. \autoref{fig:study_example} shows the snapshots of an example video generated by a participant with \tool{}. 

Then, we collected the participants' subjective ratings in the form of a 10-question survey and qualitative feedback for \tool{} from the semi-structured interview. \autoref{fig:likert_result} presents the questions and average user ratings. 
Generally, all participants agree that \tool{} is a useful tool to create expressive geographic data videos with intuitive guidance and is easy to learn.
They show a strong willingness to use \tool{} to simplify the prototyping process.

\textbf{Usability.}
All participants agreed that our tool eases manual efforts in the geographic data video creation process. 
P4 mentioned that \says{It does not require writing code or setting complex parameters.}
Our tool provides a code-free procedure to allow video creators to be more productive. Furthermore, the participants reported the effectiveness of the automatic parameter setting for the camera movements.
P1 stated that \says{Taking five minutes to generate a one-minute video is very efficient. It can save a lot of time.}
P5 described the tool's benefit for nonprofessionals: \says{I feel the tool is efficient for the nonprofessionals given the story to be told.}

Two participants showed their preference of the Timeline Panel.
P1 described how the timeline provided a general understanding of the video's structure: \says{The timeline can show the distribution of the cameras for telling a story. Thus, I can have a general idea of the video's purpose.}
P5 mentioned, \says{The timeline control is easy to use because it follows the design of regular video editing software.}

\textbf{Learnability.}
Most of our participants agreed that learning and using the system and the integrated functions for camera movements are easy. 
P1 responded very positively to learning our tool as a novice user: \says{I can quickly learn the visual effects of different camera movements after exploring the tool.} 
We also heard some descriptions of how our tool satisfies their needs to create camera movements with automatic parameter settings and self-controlled functions.
P7 commented that \says{The recommended parameters become more useful when the complexity of the camera movement increases.}
P8 also stated that \says{The system provides the user with automation and control at the same time.}
For the whole creation process with the tool, two participants (P4 and P8) suggest that our system ensures a smooth flow in creating and editing operations in \tool{}: \says{The overall (camera movement creation) process is smooth and natural.} 

\textbf{Expressiveness.}
Overall, the participants agreed on the expressiveness of our tool. They indicated that the proposed design space of the camera movements in geographic data videos satisfies their requirements for crafting videos. The participants also suggested that \tool{} could help discover new camera movements and learn the narrative purpose that the camera movements could serve.

First, all participants suggested that our design space for camera movements in geographic data videos is \says{clear and easy to understand}.
P4 stated that \says{Novice user may not know where to start to generate different camera effects for geographic data visualization.}
They described how the design space helped them create geographic data videos.
P6 noted that our design space \says{makes sense} and is \says{reasonable to describe from the what, why, and how perspectives.}
In particular, the narrative purposes in our tool can provide guidance to obtain a \says{more logical} video structure.
P8 emphasized the usage of narrative purposes: \says{The narrative purposes basically satisfy my intent to tell stories through camera shots.}

Second, our participants expressed their satisfaction with the diversity of the camera movements generated by our system. 
P1 stated that \says{The current camera effects are close to the common camera movements, including those in the data videos.} 
We observed that some participants attempted different camera shots under a specific narrative purpose prior to the final decision, such as P6: \says{I did not know many of the camera shots before. The system organizes the camera shots by purpose and teaches me how to use them.} 
P5 mentioned the final performance of the crafted videos and agreed that the final video is \says{good enough for general purposes, such as reporting findings and showing insights.} 
P8 showed great interest in \says{some unrealized camera shots} suggested in the left panel. He commented that the camera shots organized for narrative purposes could increase creativity in the video authoring process.

\textbf{Flexibility.}
Many participants appreciated the flexibility of our tool for editing camera movements.
P3 agreed that our tool provides sufficient degree of freedom to create camera movements: \says{The tool with camera module I previously used only provides limited choices of camera and usually fixed path.}
Another participant (P1) also likes the user interface for camera movement editing: \says{When editing the initial and final states, the camera movements can be accurately controlled.}
P7 liked the exporting and importing functions in the tool because they \says{make the camera movements reusable}.

\textbf{Mixed-initiative Authoring.}
In the survey, we found that many participants preferred the default mode with adaptive camera movement settings. 
In the interview session, we also asked the participants about their preference for the default mode with adaptive parameter settings and the manual mode for crafting camera movements. 
The participants stated that \says{the adaptive camera movements are convenient} (P2) and that they would love to \says{use auto mode first and then fine-tune the details} (P4).
However, we also discover that the adaptive parameter setting is not always the first option, especially for proficient users. 
P7 preferred to edit camera movements manually: \says{For simple camera movements, I would use the manual mode.} As a proficient user, P8 commented that \says{When I already know what effect I want to achieve, I use the manual mode more frequently.}

\textbf{Future Usage.}
In our interview, the participants expressed their strong willingness to use the tool in the future. 
P1 said \says{This technique can be used to generate highlight replays in e-sports. It greatly reduce efforts to show the exciting moments in a virtual space.}
P4 said \says{After modeling and editing in a 3D scene, the method can simplify the demonstration process.}
P8 has experiences in drone photography, and he mentioned that \says{Shooting with a drone is not easy. Maybe I could use this tool for planning and previewing before shooting from my drone.}
The participants also provided suggestions for future system improvement from visual design and camera creation. 
P5 noted that \says{The annotation could be better by matching the length of the text and the duration of the camera movements.}
P3 said \says{Combining different camera movements by the dragging and dropping interactions would be helpful.}
P2 and P3 voiced their confusion about the camera's moving trajectory: \says{It would be clearer to have an overview of the camera's moving trajectory in the visualization.}
P4 stated that the camera system could be integrated into a real-time monitoring system for \says{...presenting anomaly information and tracking.} 
P7 suggested \says{building a fully automatic system from detecting the insights and creating a fast-preview.} 
P2 inspired us to improve camera movements by \says{considering and optimizing the overall speed.} 
P3 and P7 suggested making the intermediate process of a camera movement more configurable (\eg, changing the path or setting an ease function for the motion) and storing the customized ones in the system. 
P6 and P7 expected that the system could consider artistic aspects of the camera effects. 
P8 suggested that \says{Whether we can change the focus of the camera and exit the current scene using blur effects.} 
The detailed comments about the user study are listed in the supplementary material.

\section{Discussion}
\label{sec:discussion}
In this section, we discuss the current limitations of our study and recommend future directions.


\textbf{Understanding the best practices in authoring geographic data video.}
We treat our design space as a probe for camera movements in geographic data videos rather than a comprehensive characterization.
First, a corpus can never be comprehensive. 
More instances may expand the sub-categories of each dimension.
Second, we only analyzed videos lasting 3-10 minutes. Therefore, the design space may not be valid for long videos or short-form videos, such as GIFs.
\rev{In addition, the camera movement recommendations in GeoCamera are based on the statistical frequencies of the combinations from the corpus. However, this approach is not always optimal for creating a compelling and persuasive camera movement given a narrative purpose. Further investigation on global optimization for camera movement configurations of different narrative purposes is promising to improve the overall engagement of the generated data videos. }
Last, when contextualizing existing taxonomies into geographic data videos, we found that not all items fit and require adaptations.
For instance, not all established camera effects were identified in the corpus, \eg, dolly zoom. 
And the ``Increasing Dynamics'' intent does not fit in the narratology.
We anticipate the future studies to evaluate and extend the design space.
For example, in-depth interviews with practitioners may reveal emerging tactics of camera movements that cover other dimensions to transit or surface geographic data insights.
As the design space categorizes general narrative purpose, geospatial target, and camera shots in geographic data stories, a closer examination of a particular category remains promising.

\textbf{Extending \tool to multifaceted authoring scenarios.}
While the usability of \tool is recognized by the target users in the user study, its design might have overlooked diversified authoring scenarios~\cite{kosara2013storytelling,Lee2015more,brehmer2021jam}.
First, our design considerations were largely based on a profile of an average user.
We strive to reduce the difficulty for amateur video makers in crafting camera movements, which is also beneficial for rapid prototyping when designing formal presentations. 
However, professionals, such as data journalists, may require more creativity support in different design phases~\cite{frich2019mapping}.
For instance, the interoperability of \tool should be improved given that practitioners often iterate between tools to achieve higher expressibility~\cite{bigelow2016iterating}.
Second, our assumption on the workflow, \ie, to connect a given sequence of story pieces covering data insights, can be overly simplified.
Prior research~\cite{Lee2015more, chen2018supporting} suggested that data storytelling is a much more complicated process, encompassing stages including exploring the dataset and selecting and organizing the findings. 
Thus, recommending data insights and story structures to alleviate human labor remains promising~\cite{chen2022does}.

\textbf{Enriching editorial layers for storytelling.}
Thus far, the research in authoring tools for data videos is still in its infancy~\cite{cao2020examining}.
We initially contributed an approach for average users to author the camera movements in geographic data videos.
Notwithstanding, many other design features also constitute a successful geographic data video.
We envision future tools to encapsulate a broader set of editorial layers for producing more engaging, persuasive, and compelling results.
For instance, visual embellishments enhance the aesthetics and imply the narrative topic~\cite{chen2022vizbelle}.
Animated narratives can better illustrate a concept~\cite{shi2021animation} or express certain emotions~\cite{lan2021kineticharts, lan2022negative}.
In terms of the cinematic effect, our current model of geographic data video supports picture-in-picture by splitting the video into halves from the middle.
The model can be further extended to enable more flexible scene arrangements.
Aligning the video with background music also remains interesting~\cite{xu2021cinematicopening}.


\section{Conclusion}
\label{sec:conclusion}

This paper presents GeoCamera, which is a geographic data video authoring tool that empowers users to tell appealing geographic stories with tailored camera movements.
Based on a \rev{design space} that summarizes diverse narrative purposes, geospatial targets, and camera shots, GeoCamera facilitates easy creation of coherent camera movements by allowing users to simply select the objects on the map and specify an appropriate narrative purpose.
GeoCamera has been evaluated with case and user studies, showing promising expressiveness and usability in helping its users to author diverse camera movements effortlessly.
In the future, we would like to extend GeoCamera to cover more complex authoring scenarios with exploration and creativity support, while providing enriched editorial layers for more engaging, persuasive, and compelling geographic storytelling.

\begin{acks}
The authors would like to thank the experts and participants for their help in the project, as well as the anonymous reviewers for their valuable comments. This work is partially supported by Hong Kong RGC GRF Grant (No. 16210321), a grant from MSRA, and NSFC (No. 62202105). 
\end{acks}

\balance

\bibliographystyle{ACM-Reference-Format}
\bibliography{reference}



\end{document}